# Coherency in One-Shot Gesture Recognition

Maria E. Cabrera, Richard Voyles, Juan P. Wachs

*Abstract*—User's intentions may be expressed through spontaneous gesturing, which have been seen only a few times or never before. Recognizing such gestures involves one shot gesture learning. While most research has focused on the recognition of the gestures itself, recently new approaches were proposed to deal with gesture perception and production as part of the same problem. The framework presented in this work focuses on learning the process that leads to gesture generation, rather than mining the gesture's associated features. This is achieved using kinematic, cognitive and biomechanic characteristics of human interaction. These factors enable the artificial production of realistic gesture samples originated from a single observation. The generated samples are then used as training sets for different state-of-the-art classifiers. Performance is obtained first, by observing the machines' gesture recognition percentages. Then, performance is computed by the human recognition from gestures performed by robots. Based on these two scenarios, a composite new metric of coherency is proposed relating to the amount of agreement between these two conditions. Experimental results provide an average recognition performance of 89.2% for the trained classifiers and 92.5% for the participants. Coherency in recognition was determined at 93.6%. While this new metric is not directly comparable to raw accuracy or other pure performance-based standard metrics, it provides a quantifier for validating how realistic the machine generated samples are and how accurate the resulting mimicry is.

## I. INTRODUCTION

The problem of recognizing gestures from a single observation is called One-Shot Gesture Recognition [1]. Current approaches to deal with this problem focus on the outcome (the sensed data associated with the gesture) rather than the process linked to the gesture generation [2]–[4]. Similarly, such approaches measure the success of one shot recognition based on the machine's ability to reach a maximum recognition accuracy, rather than comparing its ability to perform as good (or bad) as humans do. What is special in one shot learning, is the limited amount of information provided by a single observation, which makes this problem ill-posed; and therefore pure machine-learning approaches are not suitable to tackle this problem –context needs to be considered in some form.

Recognizing gestures is difficult for two reasons: gestures are intrinsically imprecise, encompassing a great deal of variability, and the humans that perform them are also imprecise, injecting characteristics of their own preferences and biomechanics. When only one example is provided this task becomes even more challenging, increasing the risk of low generalization capabilities [5]. The method presented here embraces these difficulties and leverages them for a beneficial outcome.

By including the human aspect within the framework, the kinematic and psycho-physical attributes of the gesture production process are used to support recognition. This approach presents a strategy to rely on psycho-physical factors to generate a dataset of realistic samples based on a single example. Using a single labeled example, multiple instances of the same class are generated synthetically, augmenting the dataset and enabling one-shot learning.

The recognition problem is framed by the idea of using the generation process for gestural instances rather than the instance itself. The proposed method captures significant variability while maintaining a model of the fundamental structure of the gesture to account for the stochastic process involved in the gesture production, associated with inherent non-linearity of human motor control [6]. We rely on global salient characteristics in a given gesture example that transcend variability due to human nature and are present in all examples of the same gesture class [7]. These characteristics are referred as the *gist* of a gesture, and are used towards an artificial and realistic gesture generation process.

The main focus of this paper is determining just how "realistic" the produced synthetic gestures are in the scope of human-robot interaction. Literature regarding brain activity shows that there are similarities in the motor cortex responses when a human observes other humans, and robots alike, perform gestures [8]. A robotic platform is used to perform these synthetic gestures in two different scenarios and determine the coherency between them.

The novelty in this paper is two-fold: (1) an innovative technique for artificial generation of human-like gestures extracted from a single example; (2) a novel metric of coherency which relates the level of agreement in gesture recognition accomplished by humans and machines.

## II. BACKGROUND

### A. Gesture Communication

Gestures are a basic form of communication between human beings. Young children use gestures to communicate before they learn how to talk [9]. Not only are the outcome and meaning of a gesture important, but also what gestures can tell us about the cognitive processes involved during gesture generation. Recent studies showed that gesturing plays a causal role in learning and that gesturing can promote learning [10]–[12].

Gestures offer a potential interface modality that includes control through symbolic commands, as for keyboards, and pointing attributes similar to those of the mouse, but in a more flexible, natural, and expressive form. Promoting forms of gesture recognition that are similar to the mechanisms existing in humans will allow a more natural communication than the existing ones.

*B. One-Shot Learning in Gesture Recognition*

An important landmark in one-shot learning applied to gestures was the Microsoft® initiative to start "ChaLearn Looking at People" Challenge in 2011 [13]. For two years a vast data set, of both development and validation batches, was used worldwide as training and testing data in the competition; the results for both years were reported in Guyon et.al. with partial success [14], [15]. A common theme in the proposed methods emphasized gesture representation as strictly machine learning and classification of observations regardless of the process involved in their generation. No relevance was mentioned towards the shape or characteristics of the human body performing the gestures.

Wan et al. extended SIFT to spatio-temporal features descriptors to build a codebook [16]. Testing videos were then processed and the codebook was applied to further classify using K-Nearest Neighbors algorithm. Their LD reached 0.18. Fanello et al. applied adaptive sparse coding to capture high-level feature patterns based on 3D Histogram of Flow (3DHOF) and Global Histogram of Oriented Gradient (GHOG), classified by linear SVM using a sliding window, reporting LD = 0.25 [3]. Wu et al. utilized both RGB and depth information from Kinect and an extended Motion History Image (MHI) representation was adopted as the motion descriptors, and maximum correlation coefficient for discriminatory method [17]. Their LD was 0.26. A different approach used was Histogram of Oriented Gradients (HOG) to describe the visual appearance of the gesture using DTW as classification method [18] with LD = 0.17.

More recent methods are described by Escalante et.al, where a 2D map of motion energy is obtained per each pair of consecutive frames in a video and then used for recognition after applying Principal Component Analysis (PCA) [1].

## III. METHODOLOGY

*A. Classical ML Problem Definition*

In the context of classical machine learning, let $\mathcal{L}$ describe a set or "lexicon" formed by $N$ gesture classes, $\mathcal{G}_i$, $\mathcal{L} = \{\mathcal{G}_1, \mathcal{G}_2, \ldots \mathcal{G}_i, \ldots \mathcal{G}_N\}$. Each gesture class is trained on a set of gesture instances $g_k^i$. In a way, the gesture class is a prototype group, and the members of that group are the instances $g_k^i \in \mathcal{G}_i$, where $k = 1, \ldots, M_i$ is the number of observations of gesture class $i$. Each gesture observation is a concatenation of trajectory points in 3D $g_k^i = \{(x_1, y_1, z_1), \ldots, (x_h, y_h, z_h)\}$, where $h$ is the total number of points within that gesture observation.

*B. One-Shot Learning Problem Definition*

The problem of one-shot learning is that the number of observations, $M_{N+1}$, for any new class of gesture, $\mathcal{G}_{N+1}$, is 1, which is insufficient for classical machine learning algorithms. Instead, we use contextual information from the lexicon, $\mathcal{L}$, which is incompatible with these ML algorithms. To reconcile this, we generate a sequence of artificial gesture instances, $\hat{g}_k^{N+1}$, where $k = 2, 3, \ldots, M_{N+1}$ and $M_{N+1}$ is now set to $M_{des}$, the desired number of instances required for training. (We include the observation in the artificial set.) These instances are generated from the sole observed instance, $g_1^{N+1}$, by extracting a set of "placeholders" or inflection points from it, which we label as $x_q^{N+1}$, where $q = 1, \ldots l$ and $l < h$. We refer to the set of placeholders from any gesture class, $\widetilde{G}_\iota$ (1), as the "gist of a gesture" of gesture class $i$ in lexicon $\mathcal{L}$.

$$\widetilde{G}_\iota = \{ \boldsymbol{x_q^i} = (x_q, y_q, z_q): \boldsymbol{x_q^i} \in g_k^i, \\ q = 1, \ldots, l, l < h \} \quad (1)$$

$$\widetilde{G}_\iota \in \widetilde{G_\mathcal{L}}, i = 1, \ldots, N + 1$$

This set of values is obtained using the function $\mathcal{M}$ (2) that maps from the gesture dimension $h$ to a reduced dimension $l$. This compact representation is then used to generate artificial gesture examples $\hat{g}_k^i$ for each $\mathcal{G}_i$. This is done through the function $\mathcal{A}$ (3), which maps from dimension $h$ to gesture dimension $l$ [7].

$$\widetilde{G_\iota} = \mathcal{M}(g_k^i), k = 1, i = 1, \ldots, N ; \\ g_k^i \in \mathbb{R}^{3 \times h} ; \widetilde{G_\iota} \in \mathbb{R}^{3 \times l} ; \quad l < h \quad (2)$$

$$\hat{g}_k^i = \mathcal{A}(\widetilde{G_\iota}), k = 1, \ldots, K ; i = 1, \ldots, N \quad (3)$$

A function $\Psi$ (4) maps gesture instances to each gesture class using the artificial examples.

$$\Psi: \hat{g}_k^i \to \mathcal{G}_i \quad (4)$$

Then for future instances $g^u$ of an unknown class the problem of one-shot gesture recognition (3-5) is defined as:

$$\text{Max } Z = \mathcal{W}\{\Psi(g^u), \mathcal{G}_i\} \\ s.t. i \leq N, i \in \mathbb{Z}^+, \mathcal{G}_i = \Psi(g_1^i), \Psi(g^u) \in \mathcal{L} \quad (5)$$

Where $\mathcal{W}$ is the selected metric function, for instance accuracy or F-Score.

In this paper the *coherency* metric is introduced to assess recognition in terms of human mimicry.

*C. Implementation Details*

The approach proposed in this paper is independent to a specific form of classification, $\Psi$. Furthermore, it is not conceived with a specific classification approach in mind. The expectation is that state-of-the-art classifiers should be selected to be trained with the artificial data sets created. This idiosyncratic approach is tested by training four different classification methods, currently used in state-of-the-art N-shot gesture recognition approaches, and adapt them to be used in one-shot gesture recognition.

*1) Classification Algorithms*

Four different classification algorithms were considered and their performances compared using the artificially generated data sets. The selected classification algorithms, namely HMM, SVM, CRF and DTW, are notorious for their recurrent use in state-of-the-art gesture recognition approaches. In the case of HMM and SVM, a one-vs-all scheme was used, while CRF and DTW provide a metric of likelihood to the predicted result after training is completed.

Each HMM is comprised by five states in a left-to-right configuration and trained using the Baum-Welch algorithm, which has been previously shown to generate promising results in hand gesture recognition [19].

For the SVM, each classifier in the one-vs-all scheme was trained using the Radial Basis Function (RBF) kernel. The library available in MATLAB® was used to implement SVM.

In the case of CRF, the training examples were encoded using the BIO scheme to determine the beginning (B), inside (I), and outside (O) of a gesture. The CRF++ toolkit was used to train and test this classification algorithm [20].

The DTW classification algorithm was implemented using the Gesture Recognition Toolkit (GRT) [21], which is a C++ machine learning library specifically designed for real-time gesture recognition.

*2) Data set: Microsoft Research MSRC-12*

This data set consists of sequences of human movements, representing 12 different iconic and metaphoric gestures related to gaming commands and interacting with a media player. The data set includes 6,244 gesture instances collected from 30 people. The files contain tracks of 20 joints estimated using the Kinect Pose Estimation pipeline [22].

A subset of this data set was selected. The number of gesture classes in the lexicon was reduced to 8. This reduction is to avoid gesture classes performing whole-body motions (like kicking or taking a bow) since the focus of this paper is on gestures performed with the upper limbs. Examples of the gestures in this lexicon are depicted in Fig. 1.

*3) Robotic enactment of artificially generated gestures*

In order to execute the artificially generated gestures using the Baxter robot, a registration and mapping process was conducted. It consisted of finding the transformation between the space where the trajectories were generated and the robot's operational space. A simple computer vision method was developed to recognize the extremities of Baxter's arms, and through tracking, estimate the trajectories that constitute the gestures. This method was developed to keep the methodology agnostic to the robot type. Alternatively, a different approach can be applied using the end-effector's position using topics and nodes from Robotic Operating System (ROS), however this is specific to the kinematic of the robot.

Baxter's gestures were detected using the following procedure: (i) add markers to the robot end effectors; (ii) segment the color using thresholding on the RGB channels of the image frame; (iii) apply morphological operators to get candidate hand regions represented by blobs; (iv) determine the center of mass for each blob $(x_i, y_i)$ and then complement the 3D representation using the depth value at that same center of mass in pixel values $(z_i)$. These coordinates represent the position $\vec{x}$ of a hand at time $i$.

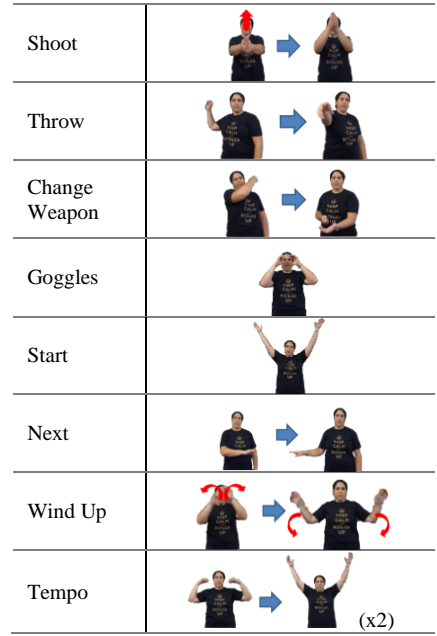

Fig. 1. Gestures selected from the MSRC-12 lexicon

### D. Performance Metrics

Once the placeholder sets, $\tilde{G}_i$, have been extracted from a single example $g_1^i$ of each gesture class $\mathcal{G}_i$ within a lexicon $\mathcal{L}$ with $i = 1, \ldots, N$, and an artificially enlarged data set has been created from it $\mathbb{G} = \{\widehat{\boldsymbol{G}}^1, \ldots, \widehat{\boldsymbol{G}}^i, \ldots, \widehat{\boldsymbol{G}}^N\}$, the goal is to evaluate the performance of the method in terms of generalization and recognition of future instances. These performance metrics involve more aspects than only accuracy. With a set of artificially generated observations for each gesture in a lexicon with their corresponding gesture class as label, each classification algorithm is trained and tested.

Confusion matrices are obtained to analyze the correlation between the actual and predicted labels of the testing data for each gesture class. The ratio between the sum of the elements in the diagonal of the matrix and the sum of all elements in the matrix is used to determine recognition accuracy.

A metric is proposed to measure the level of coherence between the recognition accuracy obtained by the proposed method, and that found when humans observe the gestures. The higher the coherence, the better the mimicry of human perception, recognition and gesture execution. This metric of coherence is related to the agreement indices $AI_x$ (6) for recognition of gestures by examining the sets of correct identifications, $C_{orr}ID_x$, and incorrect identifications, $I_{ncorr}ID_x$, for each algorithmic approach and for human participants.

$$AI_x = \frac{|C_{orr}ID_x|(|C_{orr}ID_x| - 1) + |I_{ncorr}ID_x|(|I_{ncorr}ID_x| - 1)}{(|C_{orr}ID_x| + |I_{ncorr}ID_x|)(|C_{orr}ID_x| + |I_{ncorr}ID_x| - 1)} \quad (6)$$

The intersection of these machine sets with the respective sets of correct and incorrect identifications for the humans gives us the coherency $\gamma_x(\cdot)$ (7) of each machine, or algorithmic approach:

$$\gamma_x = \frac{|C_{orr}ID_x \cap C_{orr}ID_{human}|}{|C_{orr}ID_{human}| + |I_{ncorr}ID_{human}|} + \frac{|I_{ncorr}ID_x \cap I_{ncorr}ID_{human}|}{|C_{orr}ID_{human}| + |I_{ncorr}ID_{human}|} \qquad (7)$$

It is important to note that when computing the coherency, the intersection of incorrect identifications is said to exist when both the machine and human make any incorrect identification. It is not necessary for the machine and human to misidentify a particular gesture instance as the same class. In a case of confusion, all that is important is that machine and human both misidentify the gesture.

*E. Experimental Design*

Ten participants were recruited and asked to watch a video of a person performing one example of each gesture class in the lexicon while showing the participant the gesture's respective label. Next, each participant observed Baxter perform a total of 16 gesture instances, two of each gesture class, in random order; and then they were asked to assign a label to each gesture instance performed by Baxter (e.g. the "Start" label was assigned to arms raised – see Fig. 1).

Once the experiment was concluded, participants were required to fill out a questionnaire inquiring about the correspondence between gestures and labels, the ease to remember the gestures, the effect of receiving a single example of each gesture class on their performance, and whether the characteristics of each gesture were maintained when the robot performed the gestures.

## IV. Results

The results for two different scenarios used to recognize gestures performed by a robotic platform are presented. The gestures are the result of an artificial generation process based on salient characteristics extracted from a single example of each gesture class as explained earlier in Section III. In the first scenario, the gestures are recognized using four different classification algorithms. In the second scenario, ten participants recognized the gestures performed by Baxter. Recognition accuracies were found for each scenario and then used to determine *coherency*. The coherency metric is used to provide a validation metric about the extent of which the artificial gestures resemble "human-like" gestures.

*A. Scenario 1: Robot performs gestures – Machine recognizes*

The gestures performed by Baxter are captured using the computer vision techniques described in the methodology section. Once the trajectories for all gesture samples are extracted, they are used as testing data for the different classification algorithms. Recognition accuracies for the 20 lexicon sets are depicted respectively in Fig. 2 – 5.

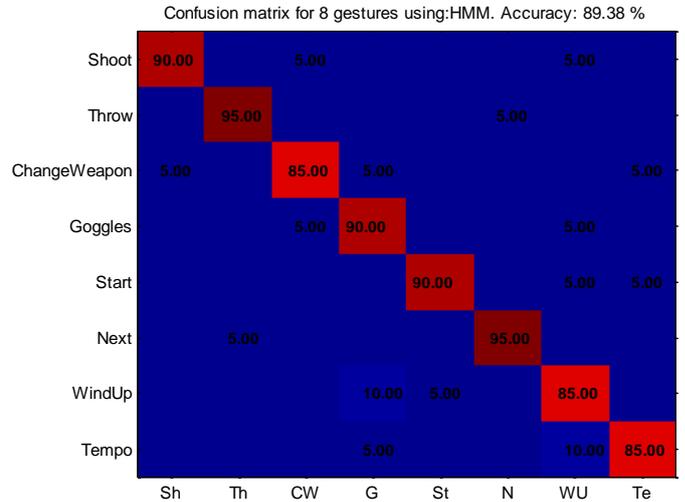
Fig. 2. Confusion matrix obtained using HMM with an accuracy of 89.38%

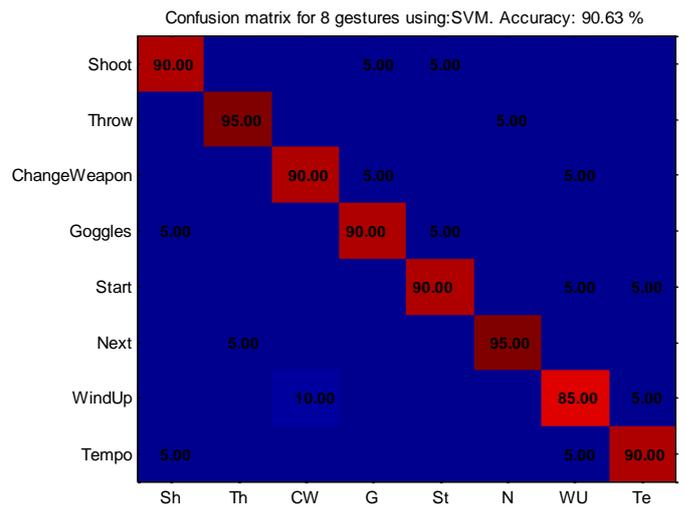
Fig. 3. Confusion matrix obtained using SVM with an accuracy of 90.63%

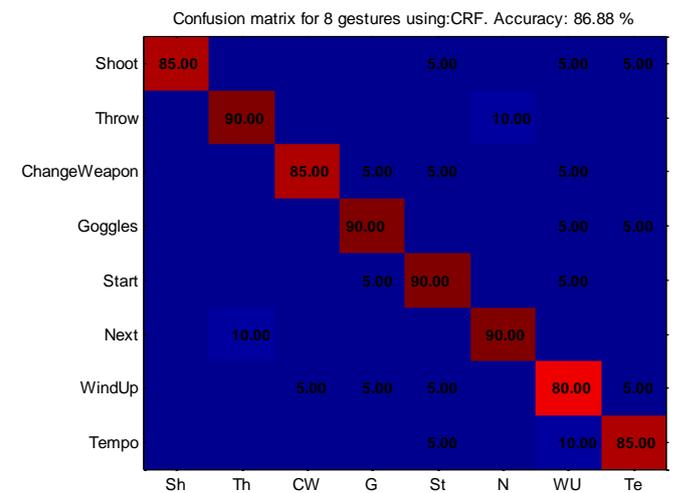
Fig. 4. Confusion matrix obtained using CRF with an accuracy of 86.88%

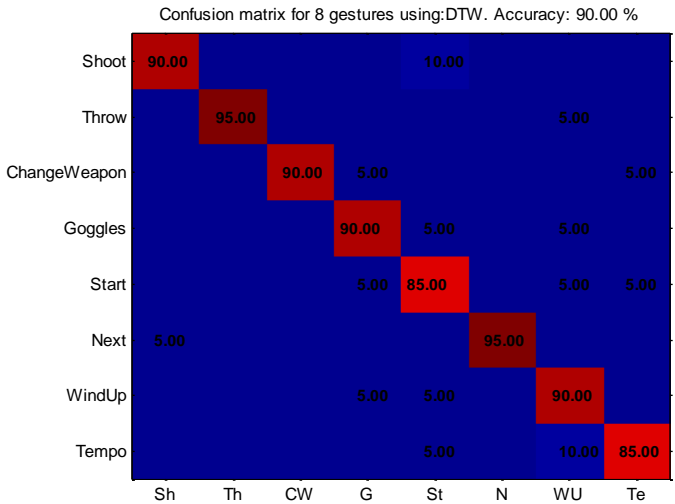

Fig. 5. Confusion matrix obtained using DTW with an accuracy of 90%

Recognition accuracy among all classifiers was specifically: HMM 89.38%, SVM 90.63%, CRF 86.88%, and DTW 90%. These results are slightly lower than state-of-the-art results found in the literature. However, the limited number of samples used may have a detrimental effect on performance when compared with standard approaches.

### B. Scenario 2: Robot performs gestures – Humans recognize

Ten participants were recruited for this experiment, 5 females and 4 males and 1 who chose not to report gender, with an average age of $26.3 \pm 3.1$ years. A permutation among the 20 artificially generated testing sets was used to assign 2 lexicon sets per participant. The order of the gesture instances performed by Baxter was randomized for each participant.

*1) Recognition Performance*

Results from the labelling process were compiled for all participants and are displayed in the confusion matrix in Fig.6. The recognition accuracy of the participants on the testing dataset was 92.5%.

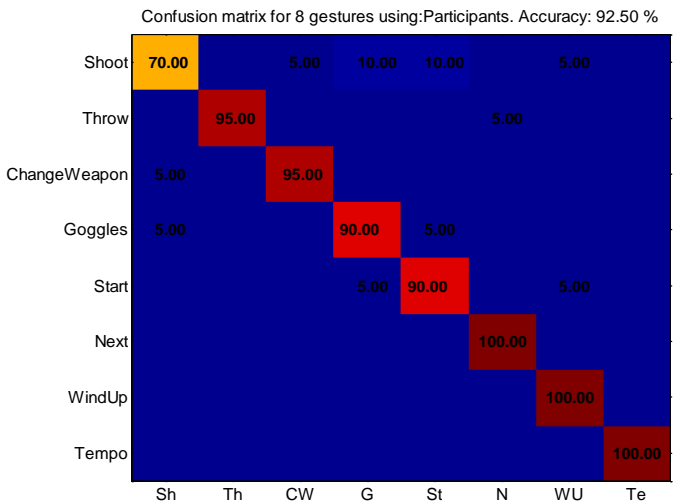

Fig. 6. Confusion matrix obtained from human participants with an accuracy of 92.5%

Three of the gestures, namely 'Next' 'Wind Up' and 'Tempo', showed the highest recognition accuracy without ever being confounded with a different gesture. Conversely, the gesture 'Shoot' showed the lowest recognition rate among the participants, getting confounded with up to four other gestures in the lexicon. One possible explanation has to do with the hand position that comes natural for humans to mimic a shooting gesture, which is very difficult to be reproduced smoothly with the robotic platform. This supports the need for including the hand configuration during the salient point extraction process.

*2) Questionnaire Results*

The answers provided by the participants in the post-experiment questionnaire are summarized in Fig.7. Aside from Likert scale type of questions, participants were required to state the frequency in which they interact with robotic platforms. The distribution of participants in frequency of interaction with robotic platforms was uniform, aiming to gather answers from a broader spectrum of potential users and not exclusively from those who may be more familiarized with robotic platforms.

The questionnaire results show agreement about the high level of correspondence between the gestures and their labels, indicating the intuitiveness of the lexicon set. For most participants, the number of gestures in the lexicon was easy to remember. However, it was noted that as the lexicon grows, the subjects' recall ability is affected. On the question related to the learning effect (examples seen before the experiment making easier to remember and recognize), participants leaned towards disagreement. This result highlights the human capability to learn from few examples. However, there was high variability in the overall response.

| Questions \ Answer Options | Strongly Agree | Agree | Undecided | Disagree | Strongly Disagree |
|---|---|---|---|---|---|
| Major correspondence between motions and labels | 4 | 6 | | | |
| Number of gestures easy to remember | 3 | 4 | 2 | 1 | |
| Number of examples seen affected performance negatively | 1 | 1 | 2 | 3 | 3 |
| Robot motions show resemblance to human motion | | 6 | 3 | 1 | |
| Robot gestures showed similar characteristics as the video | 2 | 7 | | 1 | |

| | Always | Very Often | Sometimes | Rarely | Never |
|---|---|---|---|---|---|
| Frequency of interaction with robotic platforms | 3 | 2 | 2 | 2 | 1 |

Fig. 7. Questionnaire Results

Regarding the resemblance of robot motions with human motion, participants mostly agreed but to a lower extent; this indicates that some of the motions performed by the robot may be lacking some of the attributes displayed by human motion. One example was mentioned earlier about hand configuration in the 'Shoot' gesture. When participants were asked if the characteristics of the gestures were still present in the robotic performance of the gesture, most participants agreed. This can be considered as a qualitative validation towards the process of extracting the gist of the gesture and using it to generate realistic artificial examples.

### C. Coherency

Using the recognition results for each classification method, coherency was computed for each of the four algorithms and reported in Table 1. While all algorithms performed similarly, the SVM approach scored best with the placeholders and variances identified. SVM also had the highest recognition accuracy, by a slight margin over DTW. We believe it may be possible to optimize coherency by tweaking the placeholders and their variances, but we leave that for a future investigation.

Three of the gestures in the lexicon showed high coherency between recognition scenarios above 98%. All agreements were higher among participants than among classification methods except for the 'Shoot' gesture, including three gestures in which human recognition showed perfect agreement. Overall coherency for the entire lexicon was determined at 93.6%. This result is a quantitative indication that the gesture generation process resembles closely human recognition.

TABLE I.   COHERENCY METRIC FOR HUMAN AND MACHINE AGREEMENT

| Gesture | $\gamma_{HMM}$ | $\gamma_{SVM}$ | $\gamma_{CRF}$ | $\gamma_{DTW}$ | $\gamma_{all}$ |
|---|---|---|---|---|---|
| Shoot | 68.4% | 68.4% | 77.4% | 67.9% | **70.1%** |
| Throw | 100.0% | 100.0% | 91.1% | 100.0% | **98%** |
| Change Weapon | 81.6% | 90.5% | 81.6% | 90.5% | **86.7%** |
| Goggles | 100.0% | 100.0% | 100.0% | 100.0% | **99.5%** |
| Start | 100.0% | 100.0% | 100.0% | 91.1% | **98.4%** |
| Next | 90.0% | 90.0% | 81.1% | 90.0% | **88%** |
| Wind Up | 72.1% | 72.1% | 63.2% | 80.5% | **72.5%** |
| Tempo | 72.1% | 80.5% | 72.1% | 72.1% | **74.9%** |
| **Lexicon Average** | **93.42%** | **95.59%** | **88.95%** | **94.54%** | **93.6%** |

## V. CONCLUSIONS

This paper introduces a new metric of coherency to the problem of one-shot gesture recognition in Human-Robot Interaction. An existing framework was used which focuses on the gesture generation process, using kinematic, cognitive and biomechanic characteristics of human interaction, to extract salient features of a gesture class from a single example, and use those to generate an enlarged artificial data set of realistic gesture samples. These artificial samples were validated using two different scenarios where a dual-arm robotic platform is used to execute the gesture trajectories. The first scenario involved the use of state-of-the-art classification methods to recognize the performed gestures. A second scenario relied on human recognition. The agreement between the different recognition methods based on machine learning, and the agreement between the recognition of ten participants were used to determine coherency. Such metric is our main indicator that the generated gestures capture human-like variations of gesture classes. Experimental results provide an average recognition performance of 89.2% for the trained classifiers and 92.5% for the participants. Coherency in recognition was determined at 93.6% in average for all classifiers. Future work includes computing coherency in the context of other approaches for artificial gesture generation, with different dual-arm robotic platforms. We also recognize that culture can have an important impact on gesture recognition and generation. This aspect is left for future study, as the human participants are assumed to be chosen from a similar cultural group.

## *Acknowledgment*

Acknowledgment of funding and support will be incorporated once the anonymized review process is completed.

## *References*

[1] H. J. Escalante, I. Guyon, V. Athitsos, P. Jangyodsuk, and J. Wan, "Principal motion components for one-shot gesture recognition," *Pattern Anal. Appl.*, pp. 1–16, May 2015.

[2] U. Mahbub, H. Imtiaz, T. Roy, M. S. Rahman, and M. A. Rahman Ahad, "A template matching approach of one-shot-learning gesture recognition," *Pat Recog. Lett.*, 34, no. 15, pp. 1780–1788, Nov. 2013.

[3] S. R. Fanello, I. Gori, G. Metta, and F. Odone, "Keep it simple and sparse: real-time action recognition," *J. Mach. Learn. Res.*, 14, no. 1, pp. 2617–2640, 2013.

[4] Y. Sabinas, E. F. Morales, and H. J. Escalante, "A One-Shot DTW-Based Method for Early Gesture Recognition," in *Progress in Patt. Recog., Image Analysis, Computer Vision, and Applications*, 2013, pp. 439–446.

[5] L. Fe-Fei, R. Fergus, and P. Perona, "A Bayesian approach to unsupervised one-shot learning of object categories," in *Ninth IEEE International Conference on Computer Vision*, 2003, pp. 1134–1141.

[6] D. M. Wolpert, "Computational approaches to motor control," *Trends Cogn. Sci.*, 1, no. 6, pp. 209–216, 1997.

[7] M. E. Cabrera and J. P. Wachs, "Embodied Gesture Learning from One-Shot," in *The 25th IEEE International Symposium on Robot and Human Interactive Communication*, 2016, pp. 1092–1097.

[8] B. A. Urgen, M. Plank, H. Ishiguro, H. Poizner, and A. P. Saygin, "EEG theta and Mu oscillations during perception of human and robot actions," *Front. Neurorobotics*, 7, p. 19, 2013.

[9] L. Acredolo and S. Goodwyn, "Baby signs: How to talk with your baby before your baby can talk," *Random House Ed.*, 2000.

[10] M. Chu and S. Kita, "The nature of gestures' beneficial role in spatial problem solving.," *J. Exp. Psychol. Gen.*, 140, no. 1, pp. 102–116, 2011.

[11] A. Segal, "Do Gestural Interfaces Promote Thinking? Embodied Interaction: Congruent Gestures and Direct Touch Promote Performance in Math," 2011.

[12] K. Muser, "Representational Gestures Reflect Conceptualization in Problem Solving," *Campbell Prize*, 2011.

[13] "ChaLearn Looking at People.". Available: http://gesture.chalearn.org/.

[14] I. Guyon, V. Athitsos, P. Jangyodsuk, B. Hamner, and H. J. Escalante, "Chalearn gesture challenge: Design and first results," in *IEEE Computer Society Conference on CVPRW*, 2012, pp. 1–6.

[15] I. Guyon, V. Athitsos, P. Jangyodsuk, H. J. Escalante, and B. Hamner, "Results and analysis of the chalearn gesture challenge 2012," in *Advances in Depth Image Analysis and Applications*, 2013, pp. 186–204.

[16] J. Wan, Q. Ruan, W. Li, and S. Deng, "One-shot learning gesture recognition from RGB-D data using bag of features," *J. Mach. Learn. Res.*, vol. 14, no. 1, pp. 2549–2582, 2013.

[17] D. Wu, F. Zhu, and L. Shao, "One shot learning gesture recognition from RGBD images," in *IEEE Computer Society Conference on Computer Vision and Pattern Recognition Workshops*, 2012, pp. 7–12.

[18] J. Konečný and M. Hagara, "One-shot-learning gesture recognition using hog-hof features," *J. Mach. Learn. Res.*, 15, no. 1, pp. 2513–2532, 2014.

[19] M. G. Jacob and J. P. Wachs, "Context-based hand gesture recognition for the operating room," *Pat Recog. Lett.*, 36, pp. 196–203, Jan. 2014.

[20] "CRF++.". Available: https://taku910.github.io/crfpp/.

[21] "Gesture Recognition Toolkit — NickGillianWiki." [Online]. Available: http://www.nickgillian.com/wiki/pmwiki.php/GRT/GestureRecognition Toolkit. [Accessed: 01-Jul-2016].

[22] S. Fothergill, H. Mentis, P. Kohli, and S. Nowozin, "Instructing people for training gestural interactive systems," in *Proceedings of the SIGCHI Human Factors in Computing Systems*, 2012, pp. 1737–1746.